\documentclass{emulateapj}

\usepackage{psfig}
\shorttitle{Observations of L1014}
\shortauthors{Young et al.}

\begin{document}


\title {\bf A ``Starless" Core that Isn't: Detection of a Source in the L1014 Dense Core with the Spitzer Space Telescope}
\author{Chadwick H. Young\altaffilmark{1},
Jes K. J{\o}rgensen\altaffilmark{2},
Yancy L. Shirley\altaffilmark{3},
Jens Kauffmann\altaffilmark{4},
Tracy Huard\altaffilmark{5},
Shih-Ping Lai\altaffilmark{6},
Chang Won Lee\altaffilmark{7},
Antonio Crapsi\altaffilmark{5},
Tyler L. Bourke\altaffilmark{8},
Cornelis P. Dullemond\altaffilmark{9},
Timothy Y.  Brooke\altaffilmark{10},
Alicia Porras\altaffilmark{5},
William Spiesman\altaffilmark{1},
Lori E. Allen\altaffilmark{5},
Geoffrey A. Blake\altaffilmark{11},
Neal J. Evans II\altaffilmark{1},
Paul M. Harvey\altaffilmark{1},
David W. Koerner\altaffilmark{12},
Lee G. Mundy\altaffilmark{6},
Phillip C. Myers\altaffilmark{5},
Deborah L. Padgett\altaffilmark{13},
Anneila I. Sargent\altaffilmark{10},
Karl R. Stapelfeldt\altaffilmark{14},
Ewine F. van Dishoeck\altaffilmark{2},
Frank Bertoldi\altaffilmark{4},
Nicholas Chapman\altaffilmark{6}, 
Lucas Cieza\altaffilmark{1},
Christopher H. DeVries\altaffilmark{5},
Naomi A. Ridge\altaffilmark{5},
Zahed Wahhaj\altaffilmark{12}}

\altaffiltext{1}{University of Texas at Austin, 1 University Station C1400, Austin, TX 78712-0259}
\altaffiltext{2}{Leiden Observatory, P.O. Box 9513, 2300 RA Leiden, Netherlands}
\altaffiltext{3}{National Radio Astronomical Observatory, P.O. Box 0, Socorro, NM 87801}
\altaffiltext{4}{Max-Planck-Institut F\"{u}r Radioastronomie (MPIfR), Bonn, Germany}
\altaffiltext{5}{Harvard-Smithsonian Center for Astrophysics, 60 Garden St. MS42, Cambridge, MA 02138}
\altaffiltext{6}{Astronomy Department, University of Maryland, College Park, MD 20742}
\altaffiltext{7}{Korea Astronomy Observatory, 61-1 Hwaam-dong Yusung-gu, Taejon 305-348, Korea}
\altaffiltext{8}{Harvard-Smithsonian Center for Astrophysics, Submillimeter Array Project, 645 N. A'ohoku Place, Hilo, HI 96720}
\altaffiltext{9}{Max Planck Institute for Astrophysics, Postfach 1317, D-85741, Garching, Germany}
\altaffiltext{10}{Division of Physics, Mathematics, \& Astronomy 105-24, California Institute of Technology, Pasadena CA 91125} 
\altaffiltext{11}{Division of Geological and Planetary Sciences 150-21, California Institute of Technology, Pasadena, CA 91125}
\altaffiltext{12}{Northern Arizona University, Department of Physics and Astronomy, Box 6010, Flagstaff, AZ 86011-6010}
\altaffiltext{13}{Spitzer Science Center, MC 220-6, Pasadena, CA 91125}
\altaffiltext{14}{MS 183-900, Jet Propulsion Laboratory, California Institute of Technology, 4800 Oak Grove Drive, Pasadena, CA 91109}

\begin{abstract}
We present observations of L1014, a dense core in the Cygnus region
previously thought to be starless, but data from the Spitzer Space
Telescope shows the presence of an embedded source.  We propose
a model for this source that includes a cold core, heated by the
interstellar radiation field, and a low-luminosity internal
source. The low luminosity of the internal source suggests a
substellar object.  If L1014 is representative, other ``starless"
cores may turn out to harbor central sources.
\end{abstract}

\section{Introduction}

Modern astronomers first recognized dark nebulae as regions of star
formation in the late nineteenth century. Molecular line studies
(e.g., Myers \& Benson 1983) showed that these dark nebulae contained
copious amounts of cold material.  Of these molecular cores, some were
detected by the InfraRed Astronomical Satellite (IRAS) and others,
so-called starless cores, were not.  With the Infrared Space
Observatory (ISO), Ward-Thompson et al. (2002) observed 18 starless
cores at 90, 170, and 200 $\mu$m.  While they mapped the emission at
170 and 200 $\mu$m, only one core was detected at 90 $\mu$m,
consistent with the idea that most of these starless cores have no
associated star more luminous than $\sim$0.1 L$_\odot$. The Spitzer Space Telescope (Werner et
al. 2004) offers unprecedented sensitivity and resolution at
wavelengths relevant to understanding the emission from luminous
objects in dense cores.  The Cores to Disks Legacy team (\it c2d\rm,
Evans et al. 2003) will observe over fifteen square degrees of large
molecular clouds and over 120 star-forming cores, of which about 75
are considered to be starless. The only starless core observed so far
is L1014, which presented a surprise.  Spitzer's observations reveal
an infrared point source in this supposedly ``starless'' core.

\section{Environment of L1014}

L1014 is a Lynds opacity class 6 dark cloud (Lynds 1962) that 
was undetected by both IRAS and MSX. Visser et al. (2002) observed this
core with the Submillimetre Common Users Bolometer Array (SCUBA), and
concluded that L1014 is a starless core based on two criteria: 1) it
was not detected by IRAS and 2) they could not find evidence for an
outflow.

While most authors have assumed a distance of 200 pc to L1014, there
is no firm evidence for this distance. The lack of foreground stars
indicates that it must be closer than 1 kpc. Because it lies in the
plane ($b = -0\fd25$) near the Cygnus arm, distance estimates to these
sources are unreliable.  We adopt a distance of 200 pc for L1014, but
we discuss the consequences of larger distances.

\section{Observations and Results}

For L1014, we present several new observations.  First, we used
Spitzer's InfraRed Array Camera (IRAC) and Multiband Imaging
Photometer for Spitzer (MIPS) to observe this core at wavelengths from
3.6 to 70 $\mu$m (Fazio et al. 2004, Rieke et al. 2004).  We also
present new data from the Max-Planck Millimeter Bolometer (MAMBO)
instrument on the IRAM 30-meter telescope and briefly discuss new
molecular line observations with the Five College Radio Astronomy
Observatory (FCRAO) and Seoul Radio Astronomy Observatory (SRAO).
Finally, we include data from the 2 Micron All-Sky Survey (2MASS) and
submillimeter observations from Visser et al. (2002).  

The core of L1014 was observed with Spitzer on 2 Dec 2003 in all four
IRAC bands (PROGID=139, AORKEY=5163776).  The observations in each
band consisted of two 30-second images, dithered by $\sim$10\arcsec,
resulting in a generally uniform sensitivity across the
5\arcmin$\times$5\arcmin\ field.  Using a modified version of
{\it{dophot}} (Schechter et al. 1993), we identified over 2900
different sources in the IRAC images, all but one of which are
consistent with being reddened background stars.  This one source was
distinct from the other objects by its IRAC colors
($[3.6]-[4.5]=1.65$, $[5.8]-[8.0]=1.65$); see Allen et al. (2004) and
Whitney et al. (2003) for a more detailed discussion of IRAC colors.
The observed fluxes were obtained by fitting interpolated, real point
spread functions (PSFs) to the detected sources.  IRAC fluxes for the
one source that is not consistent with being a reddened background
star, SSTc2d 2124075+495909, at RA=21$^h$24$^m$07.51$^s$ and
Dec=$+$49$^\circ$59\arcmin 09.0\arcsec\ (J2000.0), are listed in Table
1.  We refer to this source as L1014-IRS and give all positions
relative to these coordinates.

\begin{figure}
\plotone{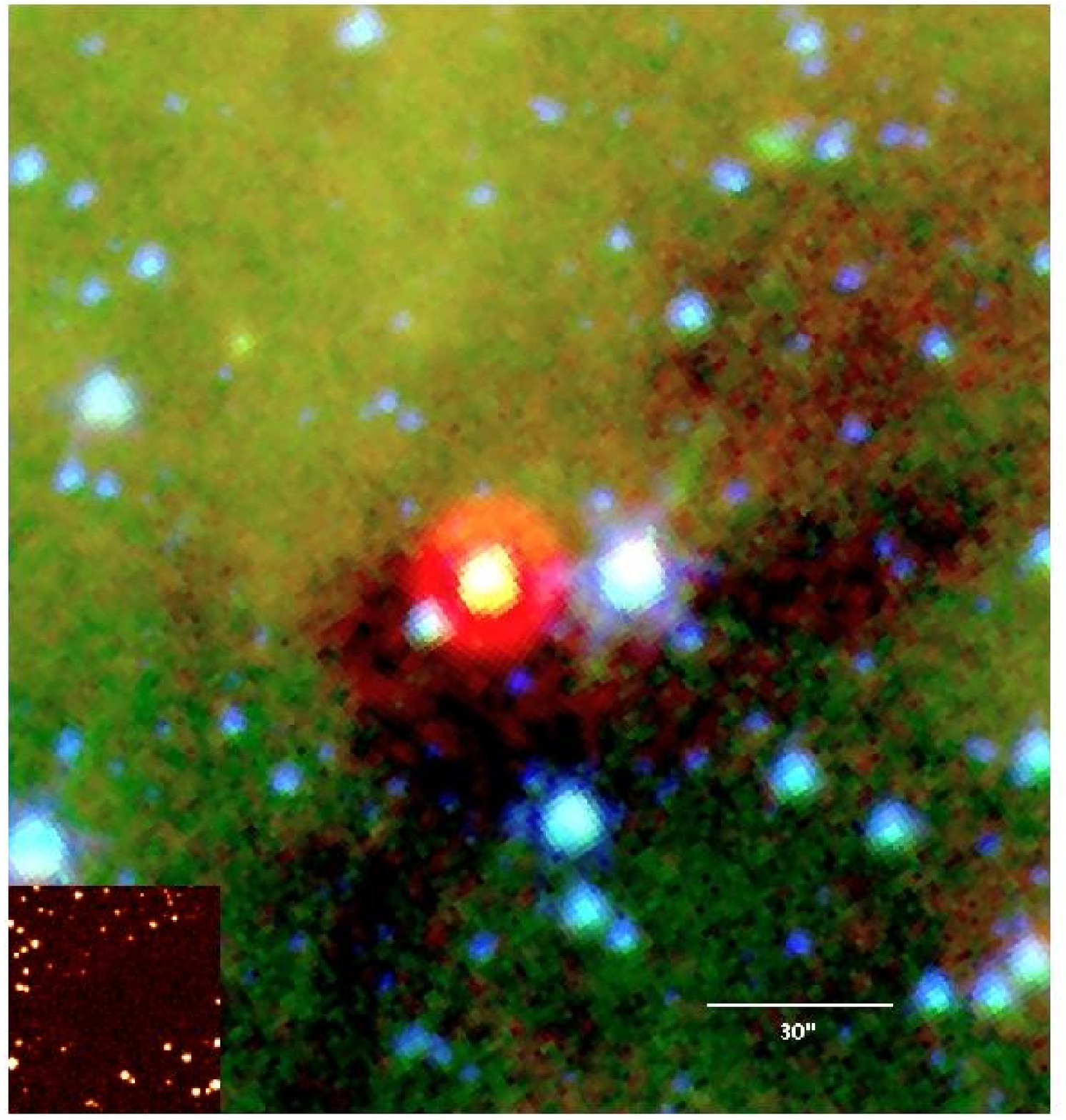} \figcaption{\label{tricolor} Three-color
composite image of L1014-IRS.  The IRAC 3.6 and 8.0 $\mu$m data are
represented here as blue and green while the MIPS 24 $\mu$m data, with
resolution more than three times worse than that of the IRAC bands,
are shown in red.  The R-band image, at lower left, is of the same
field from the Digitized Sky Survey (DSS-2). The scale and orientation
are the same as in Figure~\ref{fig2}.}
\end{figure}

MIPS 24 and 70 $\mu$m large field photometry of L1014-IRS was done on
10 Dec 2003 (PROGID=139, AORKEY=5164032).  The 24 and 70 $\mu$m
integration times were 36 and 100 seconds,respectively.  The sky
offset position was 300\arcsec\ in the scan direction (PA$\sim
45^\circ$).  The 24 $\mu$m source is shifted from the IRAC position
for L1014-IRS by ($\Delta$RA, $\Delta$Dec)=($-$0\farcs5,
0\farcs7). The 70 $\mu$m source is shifted by (1\farcs8, 1\arcsec).
The fluxes at 24 and 70 $\mu$m are in Table 1, listed with 20\%
uncertainty based on the calibration uncertainty as given by the SSC.
We show a 3-color image, including the 3.6, 8.0, and 24 $\mu$m
emission, in Figure~\ref{tricolor}. L1014-IRS stands out as the large
red source at the center.  The two closest, bright, bluer sources are
seen in 2MASS; the brighter of the two is saturated in some of the
IRAC bands.  The available data for both sources are consistent with
the idea that they are heavily reddened background stars, while the
colors of L1014-IRS are inconsistent with those of a reddened
background star.
                  
The 2MASS point source catalog does not offer measured JHK fluxes for
L1014-IRS.  However, inspection of the K-band ``Quicklook'' image
suggests the presence of faint, extended emission coincident with the
central position of L1014-IRS.  We have obtained the uncompressed
2MASS image and derived a K-band flux, in Table 1, for an 8\arcsec\
aperture.  This aperture includes all of the extended emission that is
visible in the 2MASS image.

\begin{figure}
\includegraphics[height=500pt]{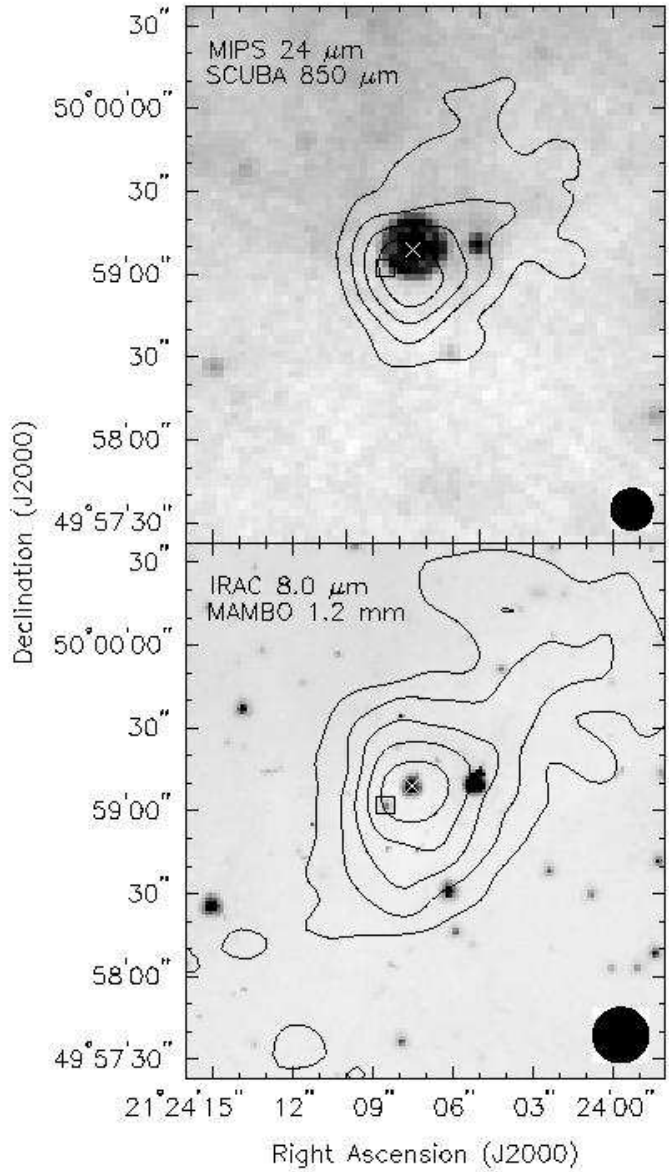}
\caption{\label{fig2} The 8.0 $\mu$m data from IRAC are shown in
greyscale with the MAMBO 1.2 mm data as contours.  The contours begin
at 3-$\sigma$ (3-$\sigma=$3.9 mJy beam$^{-1}$) and increase by
3-$\sigma$.  The white cross marks the position of L1014-IRS
(21h24m07.51s, $+$49d59m09.0s J2000).  The data has been smoothed by
20\arcsec, as shown by the black circle. The 24 $\mu$m data from MIPS
are shown in greyscale with the SCUBA 850 $\mu$m data as contours.
The contours begin at 3-$\sigma$ and increase by 3-$\sigma$.  The
black circle shows the resolution of the James Clerk Maxwell Telescope
at 850 $\mu$m, 15\arcsec. The white cross marks the position of
L1014-IRS (21h24m07.51s, $+$49d59m09.0s J2000).  Additionally, 2MASS
21240852+4959021, as discussed in the text, is marked by a white
square. }
\end{figure}

We observed the 1.2 mm continuum emission, shown overlaid on the 8.0
$\mu$m data in Figure~\ref{fig2}, in October 2002 using the 37-channel
MAMBO array (Kreysa et al. 1999) at the IRAM 30-meter telescope on
Pico Veleta (Spain). The beam size on the sky was 10\farcs7. The
source was observed on-the-fly, with the telescope secondary chopping
in azimuth by 60\arcsec\ at a rate of 2 Hz; the total on-target
observing time was 85 minutes. Zenith optical depth was 0.2.  The data
were reconstructed and analyzed using iterative methods (Kauffmann et
al., in prep.).  The map was smoothed to 20\arcsec.  The observed flux
for L1014, in Table 1, is calculated within a 120\arcsec\ aperture
centered on the position of a Gaussian fit to the data. The center of
the Gaussian is shifted from the IRAC position for L1014-IRS by
($-$1\farcs5, $-$5\farcs4).

Finally, we have collected the SCUBA data (850 and 450 $\mu$m) of
Visser et al. (2002) from the telescope's archive; the 850 $\mu$m data
are shown as contours overlaid on the MIPS 24 $\mu$m image in
Figure~\ref{fig2}.  To provide consistency with our measured flux from
MAMBO, we have taken the SCUBA fluxes with a 120\arcsec\ aperture
centered on the position of a Gaussian fit to the data.  The 850 and
450 $\mu$m peaks are shifted from the IRAC position by ($-$1\farcs5,
$-8\arcsec$) and (0\arcsec\ , $-$6\farcs9), respectively.  With these
continuum data (from $\lambda=$ 3.6 to 1200 $\mu$m), we calculate
$T_{bol}=50$ K (Myers \& Ladd 1993) and $L_{bol}/L_{smm}=20$
(Andr\'{e} et al. 1993), classifying L1014-IRS as a Class 0 core.

We observed the emission from various molecular transitions with the
SRAO 6-m telescope at Seoul National University (Koo et al. 2003),
with a beam size of $\sim 100\arcsec$.  Our CO (1$\rightarrow$0) maps
covered 480\arcsec$\times$480\arcsec\ with a 4\arcmin\ grid spacing;
the $^{13}$CO (1$\rightarrow$0) map covered 120\arcsec\ by 120\arcsec\
on a 2\arcmin\ grid.  We also made single pointing observations toward
the center of L1014 in CS (2$\rightarrow$1) and HCN (1$\rightarrow$0).
The observations of CO show that there are two different velocity
components at about 4 and $-40$ km s$^{-1}$ towards L1014 ($\sigma
=0.05$ K).  The 4 km s$^{-1}$ component was detected in CS
(2$\rightarrow$1) (T$_A^\ast\sim$0.22 K) and marginally in HCN
(1$\rightarrow$0) (0.09 K). The $-40$ km s$^{-1}$ component was not
detected in either CS or HCN ($\sigma =0.03$ K).

We also mapped L1014 in CO and C$^{18}$O (1$\rightarrow$0) using the
SEQUOIA 32-pixel, focal-plane array mounted on the 14-m FCRAO, with a
beamsize of $\sim 40 \arcsec$.  The spectra were taken in a
800\arcsec$\times$800\arcsec\ region centered at the 1.2 mm emission
peak in on-the-fly mode.  We detected strong emission of CO
(1$\rightarrow$0) ($\sigma =0.100$ K; $\Delta v=0.127$ km s$^{-1}$)
and C$^{18}$O (1$\rightarrow$0) ($\sigma =0.022$ K; $\Delta v=0.133$
km s$^{-1}$) at L1014's velocity (4.2 km s$^{-1}$); this emission
extended across our entire map.  The spectra were highly non-Gaussian
and showed a blue wing-like emission throughout the map. The CO
emission also has an additional overlapping component redward of the
C$^{18}$O line center position (at 5 km s$^{-1}$) that peaks at
($-$200\arcsec, $-$200\arcsec) and is elongated in the SW-NE
direction.  The C$^{18}$O (1$\rightarrow$0) integrated intensity peaks
at ($-$27\arcsec, 26\arcsec) from the reference position; the emission
is elongated in the SE-NW direction and is approximately 200\arcsec\
wide. The component at $v_{LSR}=-40$ km s$^{-1}$ peaks at (32\arcsec,
$-$73\arcsec) relative to our reference position. From the FCRAO and
SRAO molecular line observations, we make two conclusions: 1) while
the CO line profiles are non-Guassian, the region is too complicated
to either confirm or deny the existence of an outflow, and 2) we see a
second velocity component in the CO and C$^{18}$O spectra at about
$-40$ km s$^{-1}$.

Brand \& Blitz (1993) assigned a distance of 2.6 kpc to an HII region,
S124, whose associated molecular cloud has $v_{lsr}=-40$ km s$^{-1}$;
S124 is offset from L1014-IRS by ($3\fd5$, $-0\fd5$).  This velocity
component was not seen in our observations of HCN (1$\rightarrow$0)
and CS (2$\rightarrow$1). Because no foreground stars are observed
within $\sim$1\arcmin\ of the core's center, the dark cloud as seen in
the DSS image cannot be at 2.6 kpc, so it must be associated with the
4 km s$^{-1}$ component.
We believe that the infrared source is also associated with the 4 km
s$^{-1}$ component, but we will consider the alternative possibility.

\section{Models for the L1014 Core}

In an attempt to understand the physical state of L1014, we have used
the data to create a working model.  We used the SCUBA 850 $\mu$m
data to determine the density profile by matching our models
with the observed intensity profile at 850 $\mu$m.  We have used a
1-dimensional radiative transfer package, \it Dusty\rm\ (Ivezi\'{c} et
al. 1999), to calculate the temperature profile and the
observed SED; \it ObsSphere\rm\ (Shirley et al. 2002) was used to
simulate the SCUBA observations, including the effects of chopping.
We used the dust opacities of Ossenkopf \& Henning (1994), which
represent grains with thin ice mantles (OH5 dust) and are appropriate
for cold, dense cores (e.g., Evans et al. 2001).

The core is heated internally by a protostar and externally by the
interstellar radiation field (ISRF).  The ISRF has been attenuated
with dust whose properties are given by Draine \& Lee (1984) with
$A_{\rm V}$=0.5.  Both a shallow power-law ($n(r)\propto r^{-1.0}$)
and a Bonnor-Ebert (BE) sphere were found to provide good fits to the
observed 850 $\mu$m brightness profile, in agreement with results from
other ``starless'' cores (e.g., Evans et al. 2001).  Because it offers
a good fit to the density structure, we use the analytical
approximation given by Tafalla et al. (2002) for an isothermal BE
sphere, $n(r) = \frac{n_c}{1 + \left( \frac{r}{r_0} \right)^{\alpha}}
\;\; $ and found a good fit for $n_c = 1.5 \times 10^5$ cm$^{-3}$; for
this central density, a BE sphere is approximated with $r_0 = 4090$ AU
and $\alpha = 2.32$.  We used inner and outer radii of 50 and 15000
AU, respectively, but these are poorly constrained.  This core has
a mass of 1.7 M$_\odot$.

The long-wavelength dust continuum emission shows that a cold, dense
core is present and external heating can explain the emission longward
of 70 \micron.  The bolometric luminosity for L1014-IRS is 0.3
L$_\odot$, typical of starless cores heated only by the ISRF (Evans et
al. 2001).  In these respects, L1014 looks like a core with no
significant internal heating.

\begin{figure}
\plotone{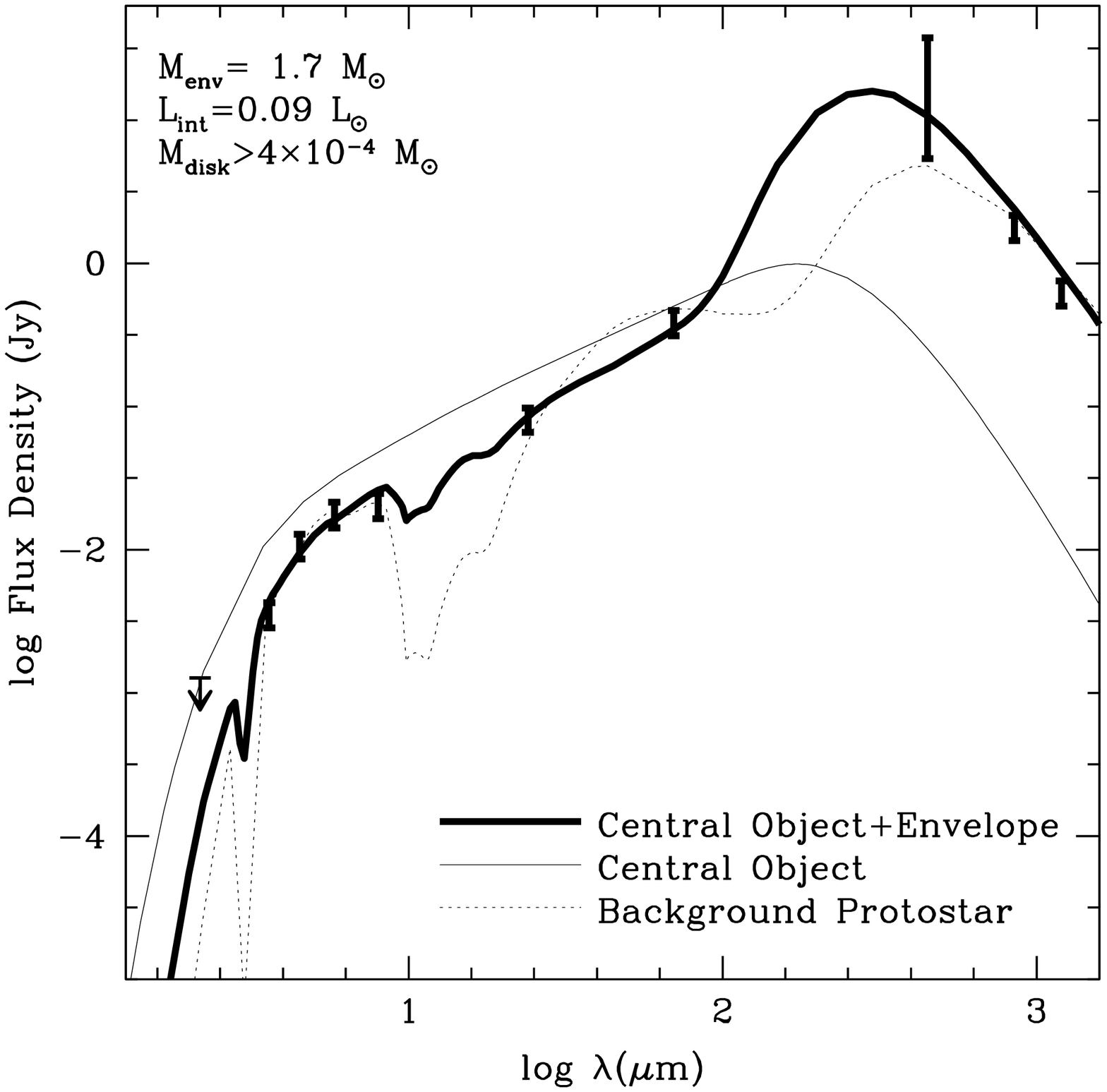} \figcaption{\label{sed}Model of the SED for L1014-IRS
assuming a distance of 200 pc.  This model shows the emission from a
Bonnor-Ebert sphere heated externally by the interstellar radiation
field and internally by a star and disk.  The star has a luminosity of
0.025 L$_\odot$ and effective temperature of 700 K.  The disk is made
to simulate flaring, has a mass of $4\times10^{-4}$ M$_\odot$, and has
an intrinsic luminosity of 0.065 L$_\odot$.  The spectrum for the
internal source is drawn by the thin, solid line; this is the input
spectrum for \it Dusty\rm.  The emergent SED is represented with the
thicker line.  The thin, dotted line shows the model for a background
protostar, as discussed in the text.}
\end{figure}

The IRAC and MIPS data trace the SED of a considerably warmer source.
The starless core model underestimates the 70 \micron\ flux by a
factor of 50 and cannot begin to explain the data at shorter
wavelengths.  Given the parameters above, we find that a
source with luminosity of 0.025 L$_\odot$ and $T_{eff}=700\pm300$ K best
fits the IRAC data (as seen in Figure~\ref{sed}).

To match the 24 and 70 $\mu$m points, a cooler component with
substantially more surface area, such as a disk, is required.  We use
a simple model (Butner et al. 1994; Adams et al 1988) to simulate the
disk emission.  The disk has a total mass (gas \& dust) of
$4\times10^{-4} (R_D/50{\rm AU})^{0.5}$ M$_\odot$, and a temperature
profile, $T(r)\propto r^{-0.5}$, simulating the effects of accretion
onto the disk surface.  $R_D$, the outer radius of the disk, is set to
50 AU, the smallest disk that matches the 70 \micron\ data; larger
values for $R_D$ and the mass are possible.

The disk model that fits is passively heated by the central protostar
and is itself luminous, possibly from accretion directly onto the
disk. The partition of luminosity between the star and disk is not
well constrained, so we use 0.09 L$_\odot$ for the total luminosity of
a central source in further discussion.  Future observations with
millimeter and submillimeter interferometers are needed to constrain
the value of $R_D$ and hence mass; this model, with $R_D = 50$ AU,
predicts a flux of 4.3 mJy at 850 $\mu$m.

If this model is correct, we are observing a very low-mass protostar
(or proto-brown dwarf).  Assuming spherical accretion at a rate of
$2\times10^{-6}$ M$_\odot$ yr$^{-1}$ (as expected for a 10 K
collapsing, isothermal sphere) onto a star with a radius of 3
R$_\odot$ and a mass of 0.1 M$_\sun$, the luminosity would be 1.8
L$_\sun$, 20 times what the observations permit. If we consider
accretion rates appropriate for the BE or ``Plummer-like'' sphere
(e.g., $3\times 10^{-5}$ M$_\odot$ yr$^{-1}$) as in Whitworth \&
Ward-Thompson (2001), the disparity between the predicted and observed
luminosity is greatly amplified.  Kenyon and Hartmann (1995) suggested
that protostars may appear sub-luminous if accretion occurs toward the
edge of a circumstellar disk. As the disk becomes too massive, the
system will undergo an FU Ori-type event where material accretes onto
the star in a short burst. Maybe, we see L1014-IRS during a quiescent
time when material is accreting and collecting in the disk.
Nonetheless, this is not an ordinary, accreting pre-main-sequence
star.  Unless accretion is much less or the radius much larger, the
mass of the central object is possibly substellar.

These conclusions depend on the distance, which is uncertain.
Roughly, the core mass and central luminosity scale as $d^2$. For
400 pc, the values would be 6.8 M$_\odot$ and 0.36 L$_\sun$.
Furthermore, the geometry might vary from the 1-dimensional core model
studied in this paper. We have created a simple 2-dimensional model
that includes a completely evacuated outflow cavity with a 30$^\circ$
opening angle and an internal source as already described.  Using
\it Radmc\rm\ (Dullemond \& Dominik, 2004), we have calculated the SED for
this model and found that, unless you look into the outflow cavity,
the SED does not change by more than 10\%.  

Another possible alternative is that L1014-IRS is a background
protostar in the $-40$ km s$^{-1}$ cloud at 2.6 kpc.  Models were
calculated for an embedded protostar at this distance obscured by a
foreground cloud; the emergent SED for this model is shown in
Figure~\ref{sed}. This was calculated in two steps in \it Dusty\rm,
first calculating the emerging spectrum of the protostar and then
illuminating a 10 K planar slab of an $A_{\rm V}=10$ with this
spectrum. The protostellar envelope was assumed to be free-falling
($n=n_0 (r/r_0)^{-3/2})$, and the ISRF was ignored.  The modeled
protostar had $L=16$ L$_\odot$ and was surrounded by a 0.6 M$_\odot$
envelope but no disk. The infrared source could be a typical protostar
at a larger distance.  We find this scenario to be statistically
unlikely for several reasons.  First, the lack of CS and HCN
emission from the $-40$ km s$^{-1}$ component and the fact that the CO
and $^{13}$CO emission do not peak at this position argue against this
scenario.  Second, a map of CS (2$\rightarrow$1) at FCRAO with a
50\arcsec\ beam shows that the 4 km s$^{-1}$ component has peak
emission within one beam of L1014-IRS, further supporting our nearby
distance (de Vries et al. in prep.).  Finally, at 70 $\mu$m, L1014-IRS
was the only source of emission within our field; the 24 $\mu$m flux
for this source was the largest within the MIPS field for L1014.  Such
bright, infrared objects are uncommon, so a chance alignment between a
background protostar and a foreground starless core is unlikely.

\section{Conclusions}

We have presented one of the most comprehensive sets of continuum data
for a ``starless'' core since their discovery.  These data span
wavelengths from 2.17 to 1200 $\mu$m and include spatially resolved
maps of the millimeter emission from L1014-IRS.  From these data, we
conclude that this core, previously thought to be starless, does
harbor a central object.  Assuming a distance of 200 pc, we have
created a model for this object that includes a 1.7 M$_\odot$ envelope
heated externally by the ISRF.  This core contains a central
object with low luminosity ($\sim 0.09$ L$_\sun$).  This central
source must have ``stellar'' and disk-like components in order to
match the 3.6 to 70 \micron\ photometry. This object appears to be
substellar, though there is enough envelope mass to grow a star if a
substantial fraction accretes.  We have considered an alternate
scenario, in which a distant, more luminous, but less embedded,
protostar lies behind the core of the L1014 dark cloud.  This scenario
is unlikely but cannot be ruled out conclusively without further
observations.

Assuming that the infrared source lies in the L1014 cloud, there are
two significant conclusions.  First, cores thought to be starless may
harbor central objects. If this phenomenon is common, estimates for
timescales for the evolution of starless cores will need revision.
Second, for plausible distances, this source has a very low
luminosity.  The usual assumptions about accretion rates and
pre-main-sequence masses and radii produce far more luminosity than is
seen. This object currently appears to be substellar, though it is
likely to accrete more mass.

Support for this work, part of the Spitzer Legacy Science Program, was
provided by NASA through Contracts Number 1224608 and 1230780 issued
by the Jet Propulsion Laboratory, California Institute of Technology
under NASA contract 1407.  The Leiden \it c2d\rm\ legacy team is
supported by a Spinoza grant from the Netherlands Foundation for
Scientific Research (NWO) and by a grant from the Netherlands Research
School for Astronomy (NOVA).  Observations at the SRAO were supported
by the KOSEF R01-2003-000-10513-0 program.  The SCUBA data were
provided to us as a guest user, Canadian Astronomy Data Centre, which
is operated by the Dominion Astrophysical Observatory for the National
Research Council of Canada's Herzberg Institute of Astrophysics. This
publication makes use of data products from the Two Micron All Sky
Survey, which is a joint project of the University of Massachusetts
and the Infrared Processing and Analysis Center/California Institute
of Technology, funded by NASA and the NSF. Finally, we thank our
anonymous referee for insightful and useful comments.

\begin{table} \label{sedtable}
\caption{Photometry of L1014-IRS}
\begin{center}
\begin{tabular}{lllll}
\hline\hline
Source & $\lambda$ & $S_\nu(\lambda)$    &  $\sigma$  & $\theta_{ap}$  \\    
  & ($\mu$m)      & (mJy) &  (mJy)  & (\arcsec) \\   
\hline\hline
L1014-IRS      				      	& 2.17        & 1.3    & 0.2   & 8    \\
$\alpha$=21$^h$ 24$^m$ 07\fs51	      	        & 3.6         & 3.6    & 0.7   & 1.7$^1$ \\
$\delta$=$+$49$^\circ$ 59\arcmin 09\farcs0  	& 4.5         & 10.7   & 2.1   & 1.5$^1$   \\
        					& 5.8         & 17.8   & 3.6   & 1.7$^1$   \\
        					& 8.0         & 20.6   & 4.1   & 2.0$^1$   \\
        					& 24          & 81.8   & 16.0  & 5.7$^1$    \\
        					& 70          & 390    & 40    & 17$^1$    \\
        					& 450         & 21500  & 16100 & 120 \\
        					& 450         & 1070   & 230   & Peak (mJy/beam) \\
        					& 850         & 1800   & 400   & 120 \\
        					& 850         & 160    & 30    & Peak (mJy/beam) \\
        					& 1200        & 630    & 126   & 120 \\
        					& 1200        & 23     & 5     & Peak (mJy/beam)\\
\hline\hline 
\multicolumn{5}{l}{$^1$FWHM of the Spitzer point spread profile. } 
\end{tabular}
\end{center}
\end{table}

\clearpage

\end{document}